# Antiferroelectric Hafnia Down to the 2D Limit

Xin Li[1]†, Guodong Ren[2]†, Haidong Lu[1], Kartik Samanta[1], Amit Kumar Shah[1], Kai Huang[1], Pravan Omprakash[2], Yu Yun[1], Pratyush Buragohain[1], Huibo Cao[3], Yan Wu[3], Jordan A. Hachtel[4], Andrew R. Lupini[4], Miaofang Chi[4], Juan Carlos Idrobo[5,6], Evgeny Y. Tsymbal[1,7]*, Alexei Gruverman[1,7]*, Rohan Mishra[2,8]*, Xiaoshan Xu[1,7]*

**Affiliations:**
[1]Department of Physics and Astronomy, University of Nebraska; Lincoln, NE 68588, USA.
[2]Institute of Materials Science and Engineering, Washington University in St. Louis; St. Louis, MO 63130, USA.
[3]Neutron Scattering Division, Oak Ridge National Lab; Oak Ridge, TN 37830, USA.
[4]Center for Nanophase Materials Sciences, Oak Ridge National Lab; Oak Ridge, TN 37830, USA.
[5]Materials Science and Engineering Department, University of Washington; Seattle, WA 98195, USA.
[6]Physical and Computational Sciences Directorate, Pacific Northwest National Laboratory; Richland, WA 99354, USA
[7]Nebraska Center for Materials and Nanoscience, University of Nebraska; Lincoln, NE 68588, USA
[8]Department of Mechanical Engineering and Materials Science; Washington University in St. Louis, St. Louis, MO 63130, USA

†These authors contributed equally: Xin Li, Guodong Ren

*Corresponding authors. Email: tsymbal@unl.edu; agruverman2@unl.edu; rmishra@wustl.edu; xiaoshan.xu@unl.edu

**Abstract:** Antiferroelectricity is a material property characterized by alternating electric dipoles spontaneously ordered in antiparallel directions. Antiferroelectrics are promising for energy storage, solid-state cooling, and memory technologies; however, these materials are scarce, and their scalability remains largely unexplored. In this work, we demonstrate that single-crystalline hafnia, a lead-free CMOS-compatible material, exhibits antiferroelectricity under compressive-strain conditions. We observe antiparallel sublattice polarization and stable double-hysteresis in single-crystalline (111)-oriented epitaxial La-doped hafnia films grown on yttrium-stabilized zirconia and show that the antipolar orthorhombic phase of hafnia adheres to the Kittel model of antiferroelectricity. Notably, compressive strain strengthens the antiferroelectric order in thinner La-doped hafnia films, achieving an unprecedented 850 °C ordering temperature in the two-dimensional limit, highlighting hafnia's potential for advanced antiferroelectric devices.

## Introduction

Antiferroelectric (AFE) crystals, as described by Kittel (*1*), consist of two sublattices, with antiparallel alignment of dipoles. Under an applied electric field, macroscopic polarization of AFE can dramatically increase due to alignment of the sublattice polarization. This field-induced large macroscopic polarization is highly desirable for low-loss, high-speed, and high-density energy storage (*2*, *3*), electrocaloric cooling, and high-speed memories (*3*).

Despite the theoretical framework, materials exhibiting antiferroelectricity, characterized by both microscopic sublattice ordering and macroscopic double-hysteresis polarization switching, remain scarce (*3*). Existing AFE materials with substantial sublattice polarization often include lead. It is theoretically predicted that at the ultrathin limit, the antiferroelectric order will diminish (*4*, *5*), although it remains largely unexplored experimentally (*3*).

Orthorhombic hafnia appears to be a promising candidate of lead-free AFE with large polarization (≈ 50 $\mu C/cm^2$) (*6–9*). Hafnia has long been used in CMOS technology as high-k dielectric materials. The recent demonstration of spontaneous macroscopic polarization in hafnia-based materials significantly boosts the feasibility of integrating ferroelectricity (FE) into modern electronics (*10*, *11*). While ferroelectric behavior in hafnia is believed to originate from an orthorhombic phase (*o-FE*, space group $Pca2_1$, **Fig. 1a**) (*12*), density functional theory (DFT) calculations indicate that a closely related antipolar orthorhombic phase (*o-AP*, space group *Pbca*, or *Pacb* if the axes align with that of $Pca2_1$, **Fig. 1b**) has a lower energy (*7*). The *o-AP* structure consists of well-separated polar layers with antiparallel polarization, matching the sublattices in Kittel's model. Although double-hysteresis-like loops have been observed in polycrystalline hafnia thin films, they are often unstable against cycling with applied electric fields (the wake-up process). Due to the coexistence of multiple phases and abundance of defects in these samples, the observed double hysteresis is attributed to defect pinning (*13–17*), ferroelastic effect (*18*), or electric-field-induced transition from the non-polar tetragonal (*t*, space group $P4_2/mnc$) phase to the *o-FE* phase (*15*, *16*, *19*, *20*). None of these mechanisms are intrinsic to Kittel AFE.

In this work, we unambiguously demonstrate the AFE state in the *o-AP* phase of hafnia evidenced by double hysteresis observation along with high-resolution scanning transmission electron microscopy and neutron diffraction measurements. Compressively strained single-crystalline epitaxial $Hf_{0.9}La_{0.1}O_2$ (LHO) films grown on (111)-oriented yttria stabilized zirconia (YSZ) exhibit an enhanced orthorhombic structural order down to one-unit-cell thickness (2D limit), which results in an unprecedentedly high Curie temperature of 850 °C.

## Single-crystalline films in the *o-AP* phase

At ambient conditions, hafnia is stable in the paraelectric monoclinic (*m*, space group $P2_1/c$) phase, followed by the *t* phase at higher temperatures (see **Fig. S1**).(*21*) Earlier theoretical works suggest that the *o-FE* and *o-AP* phases are energetically more favorable than the *m* phase in (111)-oriented hafnia films under large biaxial compressive strains (*22*, *23*). To maximize the compressive strain exerted on the hafnia film by the YSZ substrate, we choose the $Hf_{0.9}La_{0.1}O_2$ composition, since the larger ionic radius of $La^{3+}$ (110 pm), compared to $Hf^{4+}$ (76 pm), is expected to expand the hafnia unit cell, which is confirmed by DFT studies (see details in **Fig. S2**).

The single-crystalline LHO films have been grown on the (111)-oriented YSZ substrates using pulsed laser deposition, followed by the structural analysis using x-ray diffraction (XRD) (more details in **Figs. S3-S8**). The orthorhombic phases (both *o-FE* and *o-AP*) feature lattice distortions $Q_a$ and $Q_b$, in which Hf atoms are displaced along the *a* and *b* axes (relative to the *t* phase), respectively, as shown in **Fig. 1c,d**. (*6*, *12*) These distortions lead to single $(010)_o$ and single $(1\bar{1}0)_o$ (subscript "o" indicates the structural coordinates of the *o-FE* phase) diffraction peaks that are shown in **Fig. 1e** (see details in **Figs. S4-S6**) for a 21-nm-thick film (*6*). These structural properties are observed for the LHO/YSZ (111) films with thickness up to at least 30 nm (see **Fig. S6**), which is the largest thickness of orthorhombic hafnia without the *m* phase reported so far.

The single-crystalline and single-phase nature of the films have been confirmed by atomic scale characterization using aberration-corrected scanning transmission electron microscopy (STEM) (more details in **Figs. S9-S17**). A representative high-angle annular dark field (HAADF) image in **Fig. 1f** shows the atomic structure of LHO when viewed along the YSZ $[1\bar{1}0]$ direction (other images are shown in **Figs. S9-S13**). A fast Fourier transform (FFT) of the LHO region (dashed box) shows a spot corresponding to the $(010)_o$ plane, which is consistent with the $Q_b$ distortion in **Fig. 1d**, and manifest as stripes in the real-space HAADF image in **Fig. 1f**. In contrast, YSZ has a uniform Zr spacing, with no visible stripes, as confirmed by the FFT pattern. The inhomogeneity of the stripe patterns in the LHO area comes from the twin domains of three structural orientations, which rotate by 120° about the [111] direction of YSZ with respect to each other, since the stripe patterns are only expected in images viewed along the $[10\bar{1}]_o$ direction.

As illustrated in **Fig. 1a,b**, the primary difference between the *o-FE* and *o-AP* phases lies in the oxygen displacements within the polar layers: the unit cell of the *o-AP* phase is essentially a 1×2×1 super cell of the *o-FE* phase. The two neighboring polar layers in the *o-AP* phase have oxygen displacements along opposite directions. Since x-ray diffraction is insensitive to the oxygen positions, it does not detect the $(0, \frac{1}{2}, 2)_o$ peak, which is allowed for the *o-AP* phase, as shown in **Fig. 1e**. In contrast, as shown in **Fig. 1g** (more details in **Fig. S18**), substantial intensity was observed around the $(0, \frac{1}{2}, 2)_o$ peak of neutron diffraction on a 30-nm-thick film due to its sensitivity to oxygen, besides the Al (111) peak from the sample environment. The ½ index is a direct evidence of the unit cell doubling along the *b* axis. Hence, the as-grown LHO/YSZ (111) films can be characterized as the *o-AP* phase over the macroscopic scale, consistent with our DFT results (see details in **Fig. S2**).

To view the *o-AP* structure and the polar domains in real space, we used STEM imaging with the film oriented along the LHO $[100]_o$ direction (see STEM observations in **Fig. S14**). An annular bright-field (ABF) STEM image, which has high sensitivity to lighter elements (*24*), is shown in **Fig. 1h**. It clearly unveils the position of the oxygen columns and is consistent with the simulated ABF images (see the structural model and the simulated image as insets in **Fig. 1h**). After determining the position of the atomic columns in **Fig. 1h**, the polar distortions (and the domain pattern) can be derived from the offset $\delta$ of oxygen columns from the centroid of the four nearest Hf atomic columns. **Figure 1i** shows that the LHO structure consists of stripes of polar regions along the *c* axis (see details in **Figs. S15-S16**). These polar stripes are separated by non-polar spacer stripes, consistent with the structure in **Fig. 1b**. Notice that the directions of the polar displacements in neighboring polar stripes (separated by spacer stripes) are mostly aligned in an antiparallel manner, with a couple of layers having parallel alignment, which can be viewed as an antiphase boundary between two antipolar domains.

### Double hysteresis in polarization switching

The polarization switching in the *o-AP* phase of LHO is measured using Pt interdigital electrodes (IDE) on top of the films (**Fig. 2a**, see more details in **Fig. S19**). The effective electrode area is calculated using the product of finger numbers, finger length, and LHO thickness, as a semi-quantitative estimation. As shown in **Fig. 2b**, the polarization switching of a 2.2 nm LHO film with IDE of 3.5 μm gap was measured at different frequencies using a bipolar triangular voltage waveform. The double-hysteresis polarization loops are consistent with the AFE state of the *o-AP* phase. These switching loops show no noticeable changes after 5000 cycles at 20 Hz switching frequency. A frequency dependence of the double hysteresis parameters also rules out the defect pinning as a possible mechanism, since the pinched loops caused by defect pinning tend to transform into FE-like hysteresis loops upon electrical cycling (*14*), which is contrary to the results in **Fig. 2b**.

Rather, an increase in the remanent polarization with the measurement frequency suggests metastability of the *o-FE* state. This is consistent with the finding from previous DFT calculations that the energy barrier for the switching between the *o-FE* and the *o-AP* phases of $HfO_2$ is much larger than the energy difference between the two phases (*8*). Hence, when the *o-AP* phase is converted to the *o-FE* phase, the high energy barrier keeps it from immediately reverting to the *o-AP* phase, resulting in non-zero remanent polarization. Furthermore, our DFT calculations suggest that the switching between the two oppositely polarized *o-FE* states occurs through stabilization of the antipolar *o-AP* phase (see more details in **Fig. S20**), consistent with the experimental observation in Fig. 2(b).

**Enhanced orthorhombic order and record-high ordering temperature in the 2D limit**

The thickness dependence of the orthorhombic structure order of the LHO films was studied using reflection high energy electron diffraction (RHEED). **Fig. 3a** shows the spectra measured with incident electron beam along the YSZ [$11\bar{2}$] direction (see images in **Fig. S21**). The intensity of the ($3\bar{3}0$) peak which is allowed in the orthorhombic phase but forbidden in the paraelectric *t* phase, is normalized using the ($4\bar{4}0$) peak which is allowed in both the phases. To quantify the orthorhombic structure order, we use the ratio $I_{(3\bar{3}0)}/I_{(4\bar{4}0)}$ as the order parameter (*25*, *26*) and plot its thickness dependence in **Fig. 3b**, which leads to two important observations. First, $I_{(3\bar{3}0)}/I_{(4\bar{4}0)}$ increases as the thickness decreases from 10 nm, which is against the expectation of weaker AFE order in lower dimensions (*4*, *5*). Second, the enhancement continues until a clear jump of $I_{(3\bar{3}0)}/I_{(4\bar{4}0)}$ is observed between 0.6 nm and 0.5 nm, where LHO becomes thinner than one unit cell. This jump is consistent with the transition of RHEED images with incident electron beam along the YSZ [$1\bar{1}0$] direction at the same thickness (**Fig. S22**), the flat film surface revealed by the Lauer oscillation in XRD (**Fig. S7**), and the atomic terraces in atomic force microscopy (AFM) images (**Fig. S23**). In addition, the double hysteresis loop of 0.6-nm thickness LHO/YSZ (111) is also confirmed in **Fig. S19b**. Hence, LHO films adopt the AFE phase down to one-unit-cell thickness, or its 2D limit.

The stronger orthorhombic order in the 2D limit is confirmed by the temperature dependence of the order parameter $I_{(3\bar{3}0)}/I_{(4\bar{4}0)}$. As shown in **Fig. 3c**, $I_{(3\bar{3}0)}/I_{(4\bar{4}0)}$ decreases as temperature increases, suggesting the transition from the orthorhombic phase to the *t* phase at high temperature. The orthorhombic phase ordering temperature is then extracted and plotted in **Fig. 3d**. For thicker (≈ 10 nm) films, the ordering temperature is about 600 °C, close to that in Y-doped hafnia films of similar thickness (*6*, *26*, *27*). The ordering temperature increases as thickness decreases, which defies conventional expectations (*4*, *5*). This trend leads to an unprecedented 2D-limit value of 850 ± 50 °C.

**Minimal interfacial reconstruction**

To understand the enhanced orthorhombic structural order of LHO in the 2D limit, we delve into the interfacial structures of the LHO/YSZ (111) films, because interfacial reconstructions (from symmetry changes to subtle structural distortions), are in general expected at material interfaces (*28–35*). In particular, for epitaxial thin films of hafnia grown on $La_{0.7}Sr_{0.3}MnO_3$, interfacial reconstruction causes non-polar structures at the interface (*33*); correspondingly, transition temperature decreases in thinner films in the ultrathin range (*36*).

**Fig. 4a** displays a HAADF image of a 10-nm-thick LHO film viewed along the $[11\bar{2}]$ direction of YSZ. A dimerization pattern due to the pairing of Hf atoms along the horizontal direction, which is caused by the $Q_a$ and $Q_b$ distortion modes illustrated in **Fig. 1c,d,** is visible. A close-up view of the interfacial area in the yellow dashed box is shown in **Fig. 4b** (see chemical information near the interface in **Fig. S17**). As indicated in the atomic structure model in **Fig. 4c**, the horizontal distance between Hf atoms splits into two values $x_{short}$ and $x_{long}$, which are visible when the structure is viewed along the $[11\bar{2}]_o$ direction. In addition, Hf atoms in the $(111)_o$ planes displace vertically, as described by $\Delta y$. From the STEM image, $\Delta x = x_{long} - x_{short}$ and $\Delta y$ are quantified and displayed in **Fig. 4d**, where the index of the topmost YSZ monolayer is 0. Remarkably, $\Delta x$ and $\Delta y$ of LHO reach their ultimate values in the first unit cell thickness (2 monolayers, 0.6 nm) from the interface. On the YSZ side, non-zero distortions are observed in the top two monolayers, as indicated in **Fig. 4d** and visible in **Fig. 4b**. In other words, while YSZ barely reduces the distortion in LHO, LHO actually induces non-zero orthorhombic distortion in the top monolayers of YSZ.

To understand the lattice distortion at the LHO/YSZ (111) interface, we performed DFT calculations (see Methods in Supplementary Materials). As shown in **Fig. 4e**, we considered a heterostructure where five-monolayer $HfO_2$ (111) was deposited on a $ZrO_2$ (111) substrate with polarization of the $HfO_2$ layers pointing to or away from the interface. After performing the structural relaxations, we found that all $HfO_2$ monolayers including those close to the interface exhibit nearly the same structural distortions as those in bulk orthorhombic $HfO_2$. At the same time, we observed that these distortions propagate into the first two monolayers of $ZrO_2$ near the interface. Since the experimentally dominated *o-AP* phase essentially represents a 1×2×1 super cell of the *o-FE* phase, we averaged the calculated distortions $\Delta x$ and $\Delta y$ over the two polarization orientations of the *o-FE* phase. The calculated distortions appear to be in excellent agreement with our experiments, as shown in **Fig. 4d** (see **Figs. S24&S25** for oxygen displacements).

## Origin of the enhanced orthorhombic order in thinner films

With the persistence of orthorhombic order down to the 2D limit explained by the minimal interfacial reconstruction, we discuss the origin of enhancement of the orthorhombic order in thinner films. For LHO/YSZ (111), if the compressive strain exerted by YSZ is large enough so that LHO reaches the bottom of the *o-AP* phase potential well (**see Fig. S2**), the *o-AP* LHO is expected to have high stability and more so in thinner films where the strain is more effective. The high stability also makes the *o-AP* phase more robust against interfacial reconstructions.

The larger compressive strain and the corresponding high stability of the *o-AP* phase in LHO/YSZ (111), can be corroborated by comparison with pristine $HfO_2$ and HZO films epitaxially grown on YSZ (111). As shown in **Fig. 4f**, $d_{(111)}$ of LHO is systematically larger than that of pristine $HfO_2$ and HZO, which means that the same YSZ (111) substrate leads to larger compressive strain of LHO than that of pristine $HfO_2$ and HZO. The much slower relaxation of $d_{(111)}$ of LHO than that of pristine $HfO_2$ and HZO in **Fig. 4f**, suggests that orthorhombic LHO/YSZ (111) is more stable compared with that of pristine $HfO_2$ and HZO. Moreover, as shown in **Fig. 3b**, the order parameter $I_{(3\bar{3}0)}/I_{(4\bar{4}0)}$ of pristine $HfO_2$ and HZO films decreases in thinner films and drops rapidly at about 1-2 nm, which is consistent with the "normal" thickness dependence of ordering temperature, i.e., reduction in thinner films, in pristine $HfO_2$ and HZO shown in **Fig. 3d**. These observations suggest a less stable orthorhombic phase in pristine $HfO_2$ and HZO, which are more susceptible to reconstruction and relaxation, as corroborated by the morphology studies on hafnia-based films of different compositions (see **Figs. S26&S27**).

## Conclusion

Results of our study unambiguously confirm the antiferroelectric nature of the *o-AP* phase of single-crystalline LHO/YSZ (111) films. This effect is attributed to strain-induced stabilization, which enhances orthorhombic structural order down to the 2D limit and yields an unprecedentedly high ordering temperature of 850 °C. These findings demonstrate that *o-AP* phase hafnia, a lead-free material, closely adheres to the Kittel model of antiferroelectricity. Its scalability down to the 2D limit, exceptionally high ordering temperature, and CMOS compatibility make single-crystalline orthorhombic hafnia a promising candidate for both fundamental AFE studies and advanced device applications.

**Acknowledgments:** We acknowledge Dr. Huafang Li and Jian Huang of Washington University for help with TEM specimen preparation.

**Funding:** Experimental work at the University of Nebraska was supported by the National Science Foundation (NSF), grant DMR- 2419172, (A.G. and X.X) and Intel Corporation (X.L, A.S., H.L, X.X., and A.G.). R.M., G.R., and P.O. acknowledge the NSF for support through grants DMR-2122070 and DMR-2145797. M.C., A.R.L. and J.A.H. are supported by the U.S. Department of Energy, Office of Science, Basic Energy Sciences, Materials Sciences and Engineering Division. The Microscopy work was conducted as part of a user project at the Center for Nanophase Materials Sciences (CNMS), which is a US Department of Energy, Office of Science User Facility at Oak Ridge National Laboratory. This work used computational resources through allocation DMR-160007 from the Advanced Cyberinfrastructure Coordination Ecosystem: Services & Support (ACCESS) program, which is supported by NSF grants 2138259, 2138286, 2138307, 2137603, and 2138296. This research used resources at the High Flux Isotope Reactor, a DOE Office of Science User Facility operated by the Oak Ridge National Laboratory. (The beam time was allocated to DEMAND on proposal number IPTS-33539.1.) JCI was supported by the U.S. Department of Energy (DOE), Office of Science (SC), Basic Energy Sciences, Material Sciences and Engineering Division, Electron and Scanning Probe Microscopies Program, FWP 83244. Research was partly performed in the Nebraska Nanoscale Facility: National Nanotechnology Coordinated Infrastructure and the Nebraska Center for Materials and Nanoscience (and/or NERCF), which are supported by the National Science Foundation under Award ECCS: 2025298, and the Nebraska Research Initiative.

**Author contributions:** The thin film synthesis and x-ray diffraction measurements were carried out by A.S., X.L., and Y.Y. under the supervision of X.X. Thickness-resolved RHEED was measured and analyzed by X.L. and X.X. Polarization switching measurements were carried out by P.B. and H.L. under the supervision of A.G. DFT calculations were conducted by K.S., K.H., P.O., and G.R. under the supervision of E.Y.T. and R.M. The neutron diffractions were carried out by X.L. and A.S. under the supervision of H.C., Y.W., and X.X. The (S)TEM experiments were conducted by G.R. under the supervision of R.M., A.R.L., J.A.H, and M.C. J.C.I. contributed with discussions of EELS and STEM imaging data analysis. The project was conceived by X.L. and X.X. X.L., G.R., H.L, R.M, E.Y.T., A.G., and X.X. co-wrote the manuscript. All the authors discussed the results and commented on the manuscript.

**Competing interests:** Authors declare that they have no competing interests.

**Data and materials availability:** All data are available in the main text or the supplementary materials."

**Supplementary Materials**

Materials and Methods

Supplementary Text

Figs. S1 to S26

References (*21*, *24*, *37–50*)

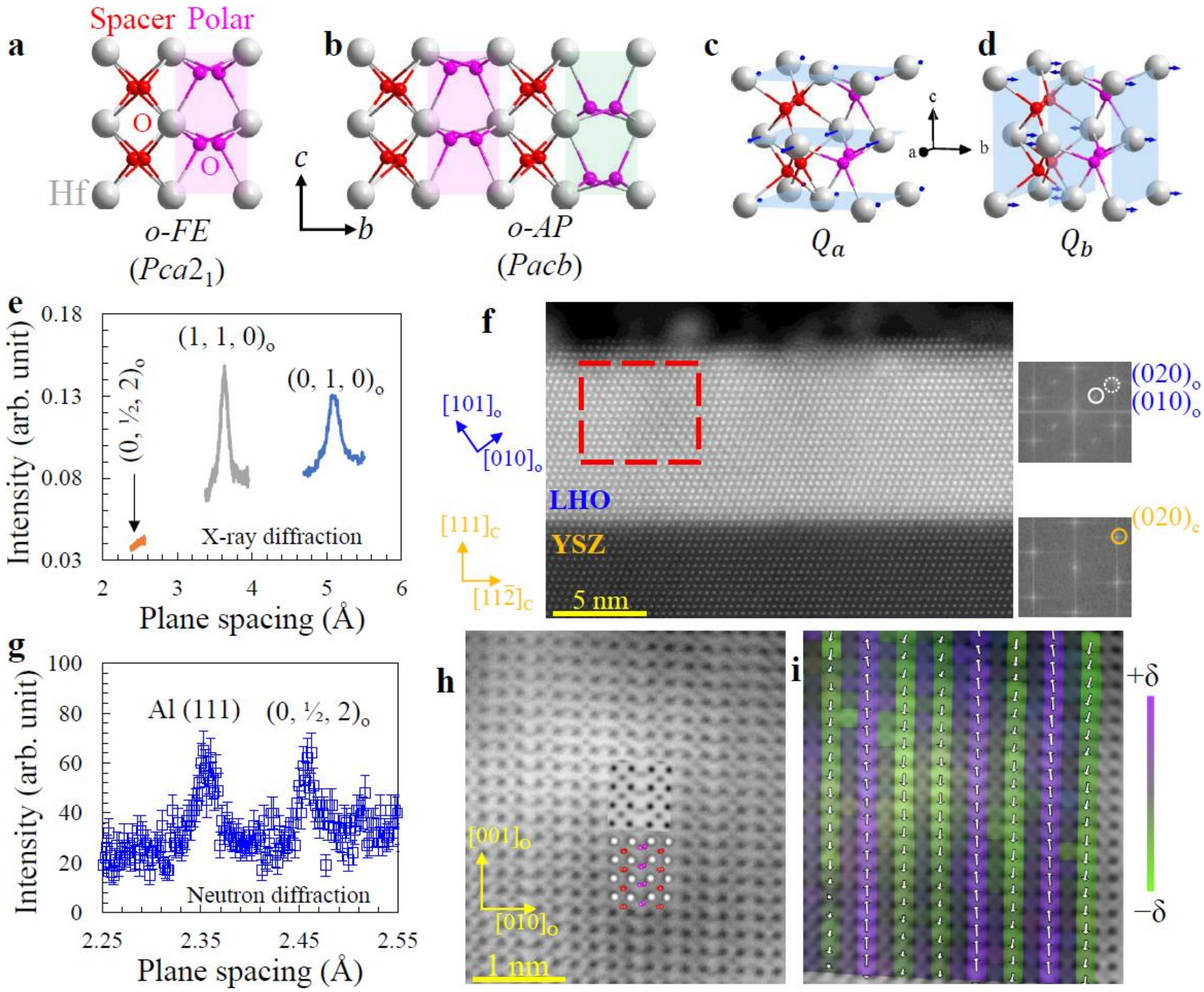

**Fig. 1. Single-crystalline epitaxial LHO/YSZ (111) films in *o-AP* phase**. (a) and (b) are crystal structure of the *o-FE* ($Pca2_1$) and *o-AP* (*Pacb*) phase of hafnia, respectively. The polar layers are indicated by the shades. (c) and (d) are the two structural distortions of the *orthorhombic* phases in terms of the Hf displacements. (e) Non-specular x-ray diffractions of a 21.4-nm LHO film. (f) A representative HAADF-STEM image viewed from the YSZ [1$\bar{1}$0] direction with fast Fourier transforms of the LHO (top right) and YSZ (bottom right) regions, respectively. (g) Non-specular neutron diffraction near $(0, ½, 2)_o$, of a 30-nm LHO film consistent with the *o-AP* phase. (h) ABF-STEM image viewed from the LHO $[100]_o$ direction, where oxygen displacements are visible. The insets are structural model (bottom) and the corresponding simulated ABF image (top). (i) Antipolar domain structure of LHO where the arrows (and color) indicate oxygen displacement δ.

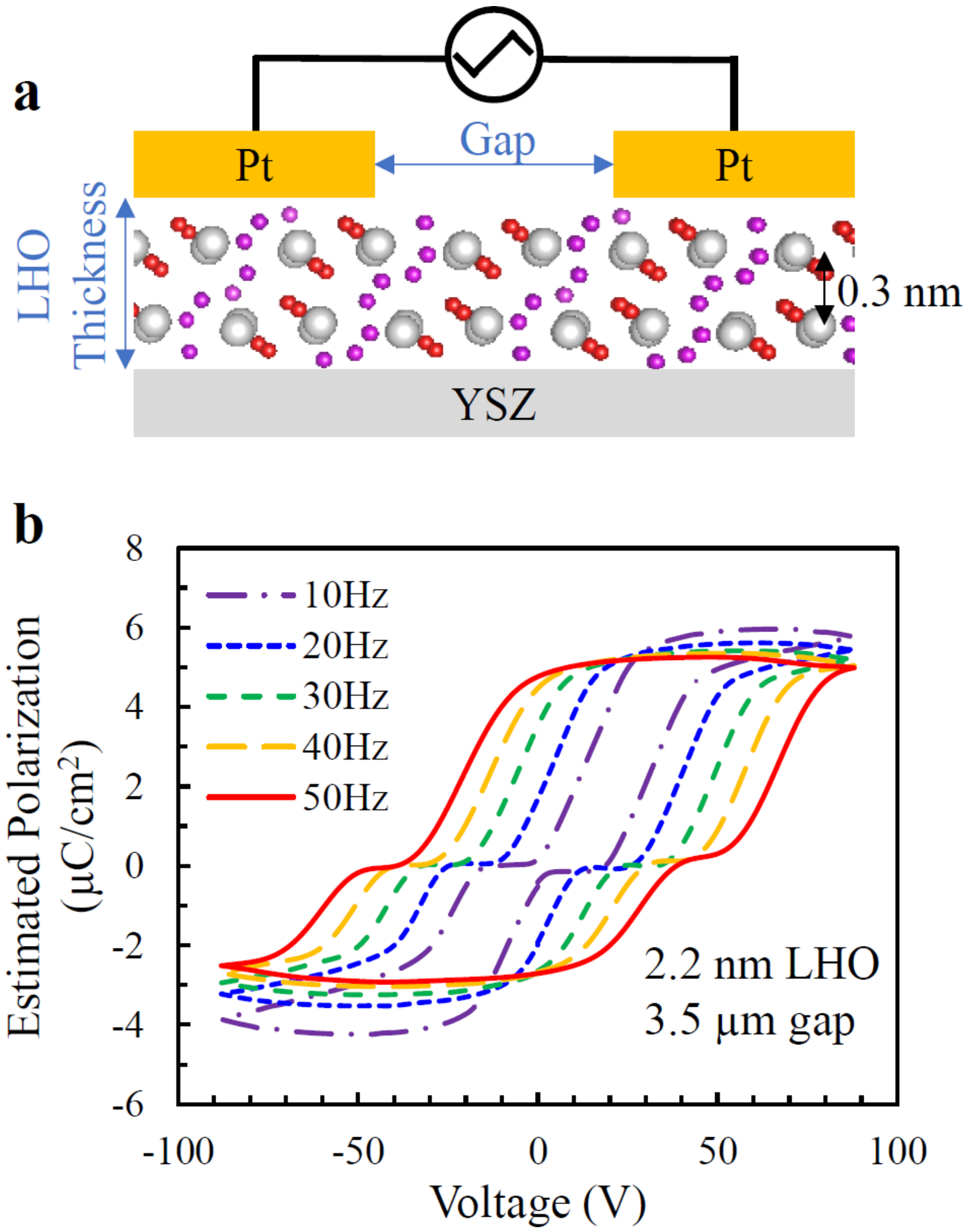

**Fig. 2. Polarization switching of LHO/YSZ (111) films**. (a) Schematic configuration of polarization switching measurements using interdigital electrodes (IDE). (b) Polarization switching hysteresis loops measured using IDE for an LHO (2.2 nm) film with different measurement frequencies.

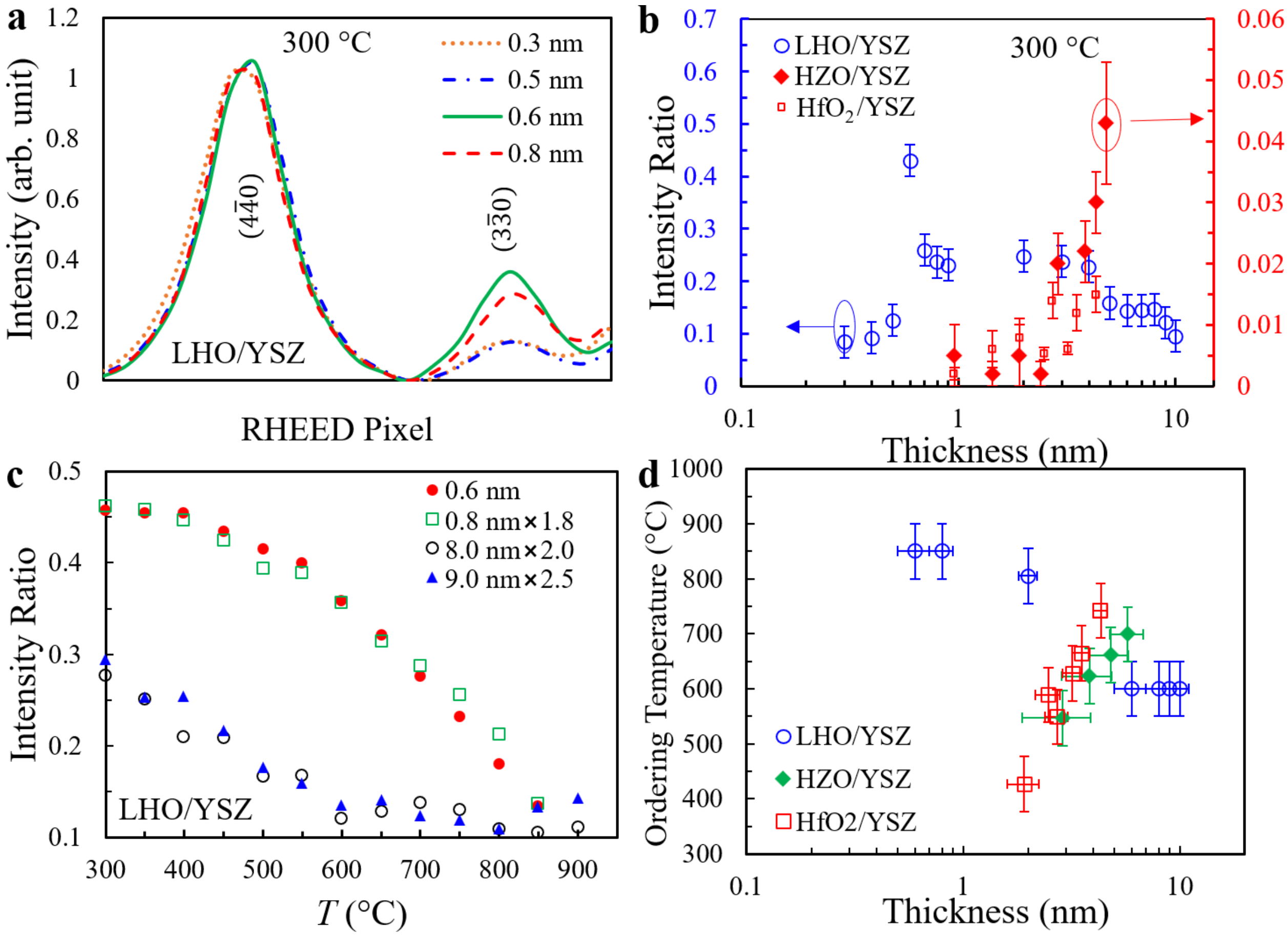

**Fig. 3. Thickness and temperature dependence of orthorhombic order of LHO/YSZ (111) and comparison with HZO/YSZ (111) and pristine $HfO_2$/YSZ (111)**. (a) Intensity profile of RHEED of different film thicknesses containing the $(3\bar{3}0)$ and $(4\bar{4}0)$ peaks with the e-beam along the $[11\bar{2}]$ direction. (b) Intensity ratio $I_{(3\bar{3}0)}/I_{(4\bar{4}0)}$ as a function of film thickness. (c) Intensity ratio $I_{(3\bar{3}0)}/I_{(4\bar{4}0)}$ as a function of temperature showing transition between the *orthorhombic* phase and the *t* phase. The data for the 0.8 nm, 8.0 nm, and 9.0 nm films are scaled. (d) Comparison of thickness dependence of the ordering temperature between different hafnia-based films.

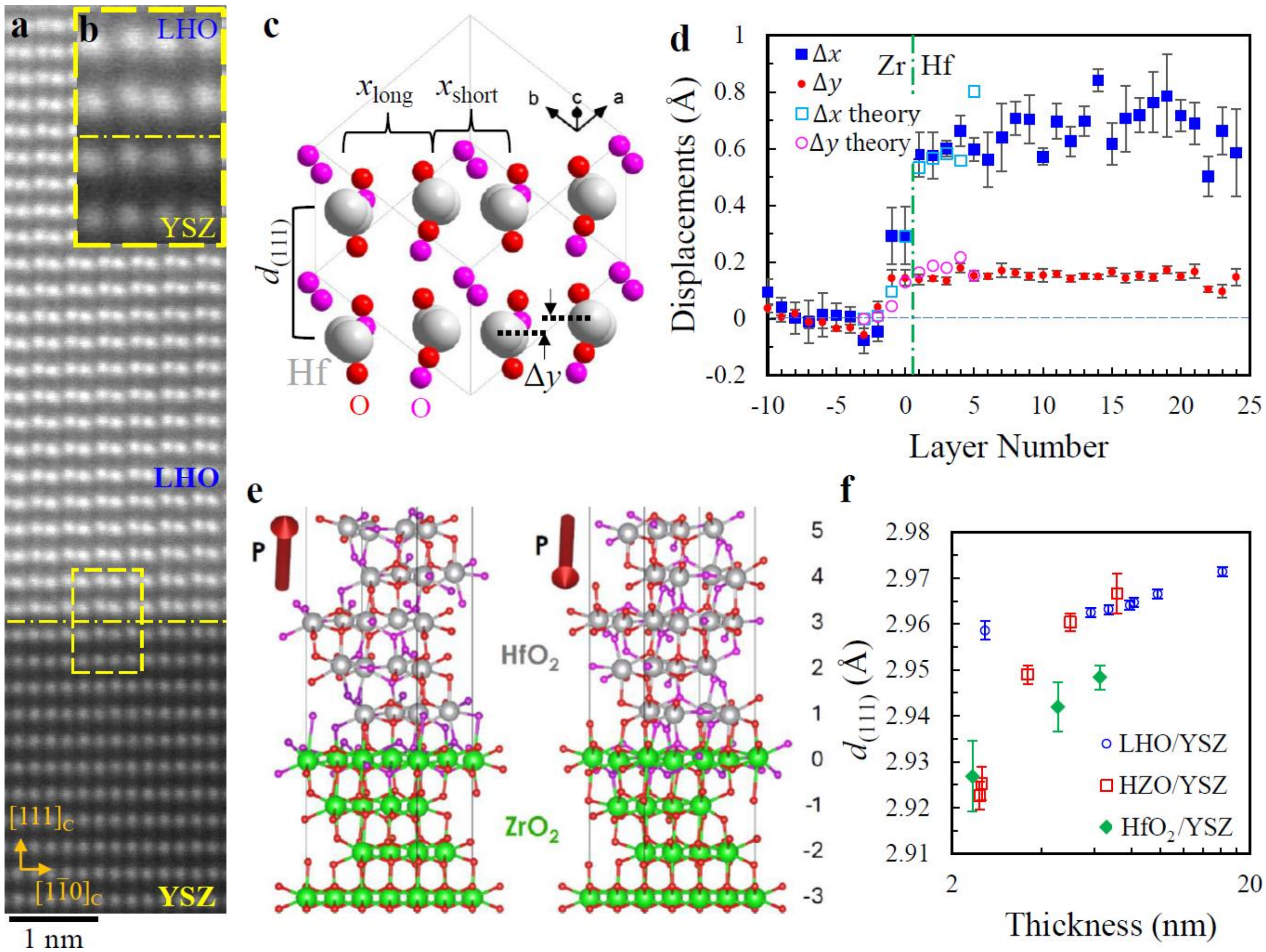

**Fig. 4. Structural distortions at the LHO (111) / YSZ (111) interface**. (a) HAADF-STEM images viewed from the YSZ [11$\bar{2}$] direction. (b) Closeup view of the dashed box in (a), where the structural dimerization is visible. (c) Structural model for the dimerization, which can be described by $x_{long}$, $x_{short}$, and $\Delta y$. (d) Experimental and theoretical values of $\Delta x \equiv x_{long}$-$x_{short}$ and $\Delta y$, which are in excellent agreement. Layer "0" is the topmost YSZ layer. (e) Interfacial structure found from DFT calculations (see text). (f) Spacing of the (111) plane as a function of film thickness measured from specular XRD on different hafnia-based films.